\documentstyle[12pt]{article}
\newcommand{\PP}{
\begin{picture}(9,3)(-2,-2)
\put(1,-12){$\tilde{}$}
\put(-2,-2){$P$}
\end{picture}}

\newcommand{\MM}{
\begin{picture}(11,3)(-2,-2)
\put(1,-12){$\tilde{}$}
\put(-2,-2){$M$}
\end{picture}}

\newcommand{\eut}{
\begin{picture}(6,3)(-2,-2)
\put(1,-14){$\tilde{}$}
\put(-2,-2){$\eta$}
\end{picture}}

\begin{document}
\baselineskip=22pt plus 0.2pt minus 0.2pt
\lineskip=22pt plus 0.2pt minus 0.2pt
\begin{center}
\vspace*{2cm}
\LARGE
The Husain-Kucha\v{r} Model:\\
Time Variables and Non-degenerate Metrics\\

\vspace*{2.5cm}

\large

J.\ Fernando\ Barbero\ G., Alfredo Tiemblo and\\
Romualdo Tresguerres

\vspace*{1.5cm}

\normalsize
Centro de F\'{\i}sica ``Miguel Catal\'an'',\\
Instituto de Matem\'aticas\\ y F\'{\i}sica
Fundamental, C.S.I.C.\\
Serrano 113 bis, 28006 Madrid, Spain
\\

\vspace{.5in}
November 26, 1997\\
\vspace{.5in}
ABSTRACT
\end{center}

We study the Husain-Kucha\v{r} model by introducing a new action principle similar to the self-
dual action used in the Ashtekar variables approach to Quantum Gravity. This new action has 
several interesting features; among them, the presence of a scalar time variable that allows the 
definition of geometric observables without adding new degrees of freedom, the appearance of a 
natural non-degenerate four-metric and the possibility of coupling ordinary matter.

\vspace*{1cm}
\noindent PACS number(s): 04.20.Cv, 04.20.Fy

\pagebreak

\setcounter{page}{1}

\section{Introduction}

In the long quest to understand General Relativity (G.R.) the use of toy models has a long 
tradition. This is especially true in Quantum Gravity and Quantum Cosmology where they have 
allowed to obtain some, otherwise very difficult to get, information. However, this does not come 
without a price because one is usually forced to introduce very strong simplifying assumptions 
and, quite often, some of the key features of the theory are lost. Though a final judgement on  
the success of this approach can only be made once a consistent Quantum Gravity theory is found, 
it is possible, in principle, to get some clues on how well one is doing by considering widely 
different toy models.  

Bianchi models (see, for example, \cite{Ryan}) are obtained by imposing homogeneity conditions on 
the gravitational variables. Their high symmetry has the consequence of killing most of the 
degrees of freedom of the full theory leaving only a finite number of them. They have been widely 
used in Quantum Cosmology mainly because the equations obtained upon quantization are  more or 
less tractable. 

There are other (less known) toy models that achieve the goal of simplifying the theory by going 
in the opposite direction: {\it adding} degrees of freedom. Chief among them is the Husain-
Kucha\v{r} model \cite{HuKu} (H-K in the following). This model is quite interesting because it 
has some of the features that make  G.R. so difficult to deal with in the quantum regime, in 
particular diffeomorphism invariance, but is significantly simpler because it lacks the 
Hamiltonian constraint (another important source of difficulties in full G.R.). This has the 
effect of {\it increasing} the number of degrees of freedom per space point from two to three.

To illustrate with a picture the different and complementary roles played by these two approaches 
one can make the following analogy: Portray G.R. as a complicated, knotted, two-dimensional-
surface $\Sigma$ embedded in ${\rm I\!R^3}$. Working with Bianchi models is something akin to 
trying to get information about $\Sigma$ by looking at a finite number of points on it. The H-K 
model, on the other hand, is like trying to gather information by studying the whole ${\rm 
I\!R^3}$. Clearly some crucial features are lost in both approaches but, still, they provide 
useful and complementary views about $\Sigma$.\footnote{\noindent This analogy is, actually, a 
little bit more than that because the Hamiltonian formulation of G.R. can be understood as the 
study, in phase space, of the hypersurface defined by the constraints.}

The H-K model, in its usual formulation (see \cite{Fer1}
for some alternative descriptions), can be conveniently derived from an action principle very 
close to the self-dual action \cite{SJS} from which the Ashtekar approach to classical and 
quantum G.R. \cite{Ash} can be found. The phase space of the Hamiltonian description of both 
theories is the same: it is coordinatized by a $SO(3)$ connection and a densitized (inverse) 
triad canonically conjugate to it. Their crucial difference is the absence of a Hamiltonian 
constraint in the H-K model. The usual interpretation of this lack of ``dynamics" is the 
following: By using the frame field in terms of which the H-K action is 
written\footnote{\noindent Being a $4\times 3$ matrix it is neither a tetrad nor a triad!} one 
can build a degenerate four-metric $g_{ab}$ and a densitized vector field $\tilde{n}^a$ (that can 
be de-densitized by means of an auxiliary space-time foliation). The lack of dynamics can be seen 
as the fact that the Lie derivative of $g_{ab}$ in the direction of $n^a$ is zero.

The four-dimensional metric that we can build from the frame field in the H-K action is 
degenerate. This can lead to the erroneous conclusion that the model describes {\it only} 
degenerate four-metrics; a fact that has induced some authors to claim, for example, that 
ordinary matter cannot be coupled to the model. We will show that this is not the case in due 
time but at this point we urge the reader to think about the following paradoxical situation: The 
fact that the Hamiltonian constraint is missing from the H-K model means that the constraint 
hypersurface of G.R. in the Ashtekar formulation is contained in the H-K one, hence, every 
solution to G.R (for example Minkowski space-time) is a solution to the H-K model. How can we 
then describe these G.R. solutions in terms of the fields present in the H-K action if we only 
have a $4\times 3$ frame field available?

The solution to this problem that we give in the paper has some unexpected implications that make 
it quite attractive. On one hand it provides an elegant way to define quantum geometric 
observables (such as areas and volumes) without having to resort to increasing the number of 
physical degrees of freedom as in previous approaches \cite{Hu},\cite{Rov1}. On the other, it 
allows the introduction of a kind of time variable in the double sense that dynamics can be 
referred to it and also that the scalar constraint (that we need now in order to get the correct 
counting of degrees of freedom) is linear in its canonically conjugate momentum (so that, upon 
quantization it gives a Schr\"{o}dinger type of equation).

The main result of the paper is that it is possible to obtain the H-K model from an action 
principle (also related to the self-dual action) that admits an interpretation in terms of non-
degenerate four dimensional metrics. This is achieved by introducing a scalar field that can be 
interpreted, in a sense that will be made more precise later, as the time variable mentioned 
before. This will not only solve the paradox presented above but also will provide a means to 
couple ordinary matter thus enhancing the usefulness of H-K as a toy model. We hope that the 
possible interpretation of this scalar field as time will help to shed some light on the problem 
of time in full G.R.

The paper is organized as follows. This introduction is followed by section II where the usual 
formulation of the Husain-Kucha\v{r} model is briefly reviewed. The new action principle, that is 
the object of this paper, is introduced in section III where we derive it from the well known 
self-dual action for G.R. The details of the Hamiltonian formulation of our model are spelled out 
in section IV. There we thoroughly study the derivation of the constraints of the theory and 
discuss their interpretation. In section V we compare the field equations in both the usual and 
the new formulation for the H-K model in order to show that they are not in contradiction (a non-
trivial fact as the number of equations is different in both cases). Section VI gives a different 
proof of the equivalence of our ``non-degenerate" formulation and the usual one at the Lagrangian 
level. We also show that the addition of a cosmological constant (made possible in our scheme by 
the availability of a non-degenerate four-metric) does not lead us beyond the H-K model. We end 
the paper with section VII, where we give our conclusions and general comments, and an appendix 
that contains some details of the computations needed to disentangle the constraints in our 
formulation. 

\section{The Husain-Kucha\v{r} Model: A Brief Review}

We review in this section the H-K model in its usual formulation in order to describe its main 
features and collect the most important formulas for future reference. We start from the action 
\cite{HuKu}
\begin{equation}
S=\frac{1}{2}\int_{{\cal M}}d^4 x \;\tilde{\eta}^{abcd}\epsilon_{ijk}e_a^i e_b^j F_{cd}^k
\label{1}
\end{equation}
where our notation is the following: ${\cal M}$ is a four-dimensional manifold ${\cal M}={\rm 
I\!R}\times\Sigma$ with $\Sigma$ a three-dimensional manifold (that we take compact and without 
boundary so that we can freely integrate by parts). Curved space-time indices are represented by 
lower case Latin letters from the beginning of the alphabet. We will make no distinction between 
4-dimensional and 3-dimensional indices. The dimensionality of a certain field will be clear from 
the context. The three and four dimensional Levi-Civita tensor densities will be denoted as 
$\tilde{\eta}^{abc}$ and $\tilde{\eta}^{abcd}$ respectively
($\eut_{abc}$ and $\eut_{abcd}$ are their inverses). We use 
the convention of representing the density weights of geometrical objects by using tildes above 
(positive) and below (negative) the stem letter representing them. Internal $SO(3)$ indices, 
running from 1 to 3 will be denoted by Latin letters form the middle of the alphabet and the 
internal Levi-Civita tensor  as $\epsilon^{ijk}$. We will also use a $SO(3)$ connection 
$A_a^i(x)$ that defines a covariant derivative acting on internal indices as 
$\nabla_a\lambda_i=\partial_a \lambda_i+\epsilon_{ijk}A_a^j \lambda^k$ and can be extended to 
space-time indices by using any torsion-free space-time connection; none of the results that we 
present in the paper will depend on the extension chosen. The curvature of $A_a^i(x)$ is defined 
as
$F_{ab}^i=2\partial_{[a} A_{b]}^i+\epsilon^{i}_{\;jk}A_a^j A_b^k$. The frame field $e_a^i$ in the 
previous action is a $4\times3$ matrix; we will reserve the name triad for its projection on the 
3-dimensional slices used in the Hamiltonian formalism.

The field equations derived from (\ref{1}) are
\begin{equation}
\begin{array}{l}
\epsilon_{ijk} e_{[b}^j F_{cd]}^k=0\\
                                   \\
\epsilon_{ijk} e_{[b}^j \nabla_{c}e_{d]}^k=0
\end{array}
\label{2}
\end{equation}
Some interesting features of (\ref{2}) are summarized in the following formulas
\begin{eqnarray}
& & {\tilde n}^a F_{ab}^i=0\nonumber\\
& & {\tilde n}^a \nabla_{[a}e_{b]}^i=0\label{3}\\
& & {\cal L}_{n^a} (e_a^i e_{bi})=0\nonumber
\end{eqnarray}
where ${\tilde n}^a=\frac{1}{3!}{\tilde\eta}^{abcd}\epsilon_{ijk}e_b^i
e_c^j e_d^k$, $n^a={\tilde n}^a/{\tilde e}$, and ${\tilde e}$ is defined by means of an auxiliary 
foliation defined by a scalar function $t$ as ${\tilde e}\equiv {\tilde n}^a \partial_a t$. 
${\cal L}_{n^a}$ denotes the Lie derivative along the direction defined by $n^a$. The first two 
equations 
in (\ref{3}) explain why we do not have a dynamics in the model \cite{HuKu} (the projections of 
the field equations onto the direction normal to the spatial slices are zero) while the last one, 
which is a consequence of the others, displays this lack of evolution as the fact that the Lie 
derivative of the degenerate four-metric $e_a^i e_{bi}$ along $n^a$ is zero.

The meaning of this model is best understood in the Hamiltonian framework. In order to define it 
we introduce a foliation by means of a scalar function $t$ and a congruence of curves (nowhere 
tangent to the surfaces of the foliation) parametrized by $t$ whose tangent vectors we denote 
$t^a$. By doing this we have that $t^a\partial_a t=1$ and, hence, the time derivatives can be 
interpreted as the Lie derivatives along the direction defined by $t^a$. We can write (\ref{1}) 
as
$$
S=\int dt\;\int_{\Sigma}d^3 x\;\left\{{\dot A}_a^i\left[{\tilde \eta}^{abc}\epsilon_{ijk}e_b^j 
e_c^k\right]+A_0^i\nabla_a\left[{\tilde \eta}^{abc}\epsilon_{ijk}e_b^j 
e_c^k\right]+e_0^i\left[{\tilde \eta}^{abc}\epsilon_{ijk}e_a^j F_{bc}^k\right] \right\}$$
where the dots denote time derivatives of the fields (Lie derivatives along the direction defined 
by $t^a$), $A_0^i\equiv t^a A_a^i$, and $e_0^i\equiv t^a e_a^i$.
After following the usual Dirac procedure \cite{Dir} one finds out that the phase space of the 
model is coordinatized by an $SO(3)$ connection $A_a^i$ and a canonically conjugate densitized 
triad ${\tilde E}^a_i$. The first class constraints are
$$
\begin{array}{l}
\nabla_{a}{\tilde E}^a_i=0\\
                          \\
{\tilde E}^a_i F_{ab}^i=0
\end{array}
$$
The first constraint (Gauss law) generates internal $SO(3)$ rotations whereas the second (known 
as vector constraint) generates spatial diffeomorphisms\footnote{\noindent Diffeomorphisms are 
actually generated by a linear combination of the Gauss law and the vector constraint.}. As we 
can see there is no scalar constraint so that we have three degrees of freedom per space point.

\section{From The Self-Dual Action to the Husain-Kucha\v{r} Model}

In this section we introduce a modified action principle for the Husain-Kucha\v{r} model that 
allows us to use four dimensional, non-degenerate metrics in order to describe it. We take as the 
starting point the self-dual action\footnote{\noindent We actually use anti-self-dual fields for 
calculational purposes.} of Samuel, Jacobson, and Smolin \cite{SJS}
\begin{equation}
S=-\frac{1}{2} \int_{\cal M}d^4 x\;{\tilde\eta}^{abcd}e_a^I e_b^JF_{cdIJ}^-
\label{6}
\end{equation}
where now $e_a^I$ is a genuine tetrad field and $I=0,\ldots, 3$ are $SO(4)$ indices, $F_{ab}^{IJ-
}$ is the curvature of an
anti-self-dual connection $A_a^{-IJ}$ defined by 
$ F_{ab}^{IJ-}=2\partial_{[a}A_{b]}^{-IJ}+2A_{[a}^{-IK}A_{b]K}^{-\;\;J}$. Following \cite{Fer2} 
we write
$$
A_a^{-IJ}\equiv \left[
\begin{array}{cc}
0 & A_a^j\\
  &   \\
-A_a^i & -\epsilon^{ijk}A_{ak}
\end{array}\right]
\hspace{3cm}e_a^I\equiv\left[-\frac{1}{2}v_a\;\;\;\;\;\;e_a^i\right]
$$
So that (\ref{6}) becomes
\begin{equation}
S=\frac{1}{2}\int_{\cal M}d^4x\;{\tilde\eta}^{abcd}\left[
v_a e_b^iF_{cdi}+\epsilon^{ijk}e_{ai}e_{bj}F_{cdk}\right]
\label{8}
\end{equation}
As we can see the (anti)-self-dual action can be obtained by adding a term involving a 1-form 
field $v_a$ to the usual Husain-Kucha\v{r} action (\ref{1}). A full discussion of (\ref{8}) can 
be found in \cite{Fer2}.

In the view of the previous formula it is natural to wonder what happens if instead of taking 
$v_a$ as a general one-form one considers it to be the gradient of a scalar $\nabla_a \phi$. Do 
we still have G.R. or something else? Let us consider then the following action
\begin{equation}
{\hat S}=\frac{1}{2}\int_{\cal M}d^4x\;{\tilde\eta}^{abcd}\left[
-e_a^iF_{bci}\nabla_d\phi +\epsilon^{ijk}e_{ai}e_{bj}F_{cdk}\right]
\label{9}
\end{equation}
Before attempting to unravel its physical meaning, some preliminary remarks are in order. First 
of all the action is no longer $SO(4)$ invariant\footnote{\noindent Because the gradient of a 
scalar function does not transform as the zero component of a $SO(4)$ vector \cite{Fer2}.} 
although it is obviously $SO(3)$ invariant. Second, we see now that $S$ is linear in the time 
derivatives of  $\phi$ so we expect to have a scalar constraint linear in its canonically 
conjugate momentum (that after quantization will lead to a Schr\"{o}dinger type of equation). It 
is natural to wonder if (\ref{9}) could be an action for gravity (with an explicit time variable 
given by the scalar field $\phi$). The answer turns out to be in the negative though, at the end 
of the day, one discovers that (\ref{9}) is still interesting in its own right. In order to check 
whether (\ref{9}) describes G.R. or not we consider the field equations coming form (\ref{6}) 
(remembering that we take now $e_a^0=\nabla_a\phi$). The field equation obtained by varying with 
respect to $A_{IJ}^-$ is
\begin{equation}
\left[\nabla_{[a}\left(e_b^I e_{c]}^J\right)\right]^-=0
\label{10}
\end{equation}
From (\ref{10}) we find out immediately that $A_{IJ}^-$ is equal to the anti-self-dual part of 
the $SO(4)$ connection $\Gamma_a^{IJ}$ compatible with $e_a^I$ defined by 
\begin{equation}
{\cal D}_a e_b^I=\partial_a e_b^I-\Gamma_{ab}^c e_c^I+\Gamma_a^{IK}e_{bK}=0
\label{11}
\end{equation}
where $\Gamma_{ab}^c$ is the Christoffel symbol of the four-metric $g_{ab}\equiv e_a^I e_{bI}$. 
Notice that, generically, the determinant of $e_a^I$
$$
\det e_a^I=\frac{1}{3!}{\tilde\eta}^{abcd}(\nabla_a\phi)
\epsilon^{ijk}e_{bi}e_{cj}e_{dk}
$$
is different from zero so that we can invert (\ref{11}) to write $\Gamma_a^{IJ}$ in terms of 
$e_{a}^{I}$ and its derivatives. By substituting $A_{a}^{-IJ}=\Gamma_{a}^{-IJ}[e,\phi]$ back in 
(\ref{6}) we get
$$
S=\int_{\cal M}d^4y\;\sqrt{g[e,\phi]}R[e,\phi]
$$
where $R$ is the scalar curvature of $g_{ab}\equiv e_a^I e_{bI}=\nabla_a\phi\nabla_b\phi+e_a^i 
e_{bi}$. 
If, by choosing $e_a^i$ and $\phi$ we can generate arbitrary and non-correlated 
$g_{ab}[e,\phi](x)$ and 
$$
\delta g_{ab}[e,\phi](x)=\int_{\cal M}d^4y\;\left[\frac{\delta g_{ab}(x)}{\delta e_c^i(y)}\delta 
e_c^i(y)+\frac{\delta g_{ab}(x)}{\delta\phi(y)}\delta\phi(y)\right]
$$ 
then $S$ must be an action for full G.R., otherwise,  it is something else. At a certain point 
with coordinates $x$ it is indeed true that both $g_{ab}$ and $\delta g_{ab}$ can be chosen to be 
anything we want. However, it is not clear that the same conclusion is true {\it for all the 
points} in a neighborhood of $x$ due to the restrictions that we have imposed to the form of some 
of the components of the tetrads (in fact the main result of the paper shows that $g_{ab}$ and 
$\delta g_{ab}$ {\it are not} completely arbitrary in all the points of $\Sigma$).

\section{Hamiltonian Formulation for the New Action}

By introducing a foliation as in section II we can write
$$
{\hat S}=\int dt\;\int_{\Sigma}d^3 x\;\left\{{\dot A}_a^i{\tilde 
\eta}^{abc}\left[\epsilon_{ijk}e_b^j e_c^k-e_{bi}\nabla_c\phi\right]+A_0^i\nabla_a\left[{\tilde 
\eta}^{abc}\left(\epsilon_{ijk}e_b^je_c^k-e_{bi}\nabla_c\phi\right)\right]+\right.
$$
\vspace*{-.5cm}
$$
\hspace{3cm}\left.+\frac{1}{2}{\dot\phi}\,{\tilde \eta}^{abc}e_a^iF_{bci}+e_0^i{\tilde 
\eta}^{abc}\left[\epsilon_{ijk}e_a^j F_{bc}^k-\frac{1}{2} F_{abi}\nabla_c\phi\right] 
\right\}\equiv\int dt\;L(t)
$$
We denote ${\tilde\pi}^a_i(x)$, ${\tilde\pi}_i(x)$, ${\tilde\sigma}^a_i(x)$, 
${\tilde\sigma}_i(x)$, and ${\tilde p}(x)$ the momenta canonically conjugate to $A_a^i(x)$, 
$A_0^i(x)$, $e_a^i(x)$, $e_0^i(x)$, and $\phi(x)$ (with Poisson brackets given symbolically by 
$\{q,p\}=1$). We find the following primary constraints
\begin{eqnarray}
& & {\tilde\pi}^a_i+{\tilde\eta}^{abc}\left[e_{bi}\nabla_c\phi-
\epsilon_{ijk}e_b^je_c^k\right]=0\label{15a}\\
& & {\tilde\pi}_i=0\label{15b}\\
& & {\tilde\sigma}^a_i=0\label{15c}\\
& & {\tilde\sigma}_i=0\label{15d}\\
& & 2{\tilde p}-{\tilde\eta}^{abc}e_a^iF_{bci}=0\label{15e}
\end{eqnarray}
The Hamiltonian and the total Hamiltonian are
\begin{eqnarray}
& &  \hspace{-1cm}
H=\int_{\Sigma}d^3x\;\left\{e_0^i{\tilde\eta}^{abc}\left[
\frac{1}{2}F_{abi}\nabla_c\phi -\epsilon_{ijk}e_a^jF_{bc}^k\right]+A_0^i\nabla_a\left[
{\tilde\eta}^{abc}\left(e_{bi}\nabla_c\phi -\epsilon_{ijk}e_b^je_c^k\right)\right]\right\}
\label{16}\\
& & \hspace{-1cm}
H_T=H+\int_{\Sigma}d^3x\;\left\{\lambda_a^i\left[
{\tilde\pi}^a_i+{\tilde\eta}^{abc}\left(e_{bi}\nabla_c\phi -\epsilon_{ijk}e_b^j 
e_c^k\right)\right]+\right.\label{17}\\
& & \hspace{5.4cm}+\lambda^i{\tilde\pi}_i+\mu_a^i{\tilde
\sigma}^a_i+\mu^i{\tilde\sigma}_i+\left.\zeta\left[2{\tilde p}-
{\tilde\eta}^{abc}e_a^iF_{bci}\right]\right\}
\nonumber
\end{eqnarray}
where $\lambda_a^i(x)$, $\lambda^i(x)$, $\mu_a^i(x)$, $\mu^i(x)$, and $\zeta(x)$ are arbitrary 
(at this stage) Lagrange multipliers. The conservation under the evolution defined by $H_T$ of 
the primary constraints (\ref{15a}-\ref{15e}) gives the following secondary constraints
\begin{eqnarray}
& & \nabla_a\left[{\tilde\eta}^{abc}\left(e_{b}^{i}\nabla_c\phi -
\epsilon^{ijk}e_{bj}e_{ck}\right)\right]=0\label{18}\\
& & {\tilde\eta}^{abc}\left[\epsilon_{ijk}e_a^jF_{bc}^k-\frac{1}{2} 
F_{abi}\nabla_c\phi\right]=0\label{19}
\end{eqnarray}
and the following conditions on the Lagrange multipliers
\begin{eqnarray}
& & \hspace{-2cm}
{\tilde\eta}^{abc}\left[\left(\frac{1}{2}
\delta_{ik}\nabla_b\phi+\epsilon_{ijk}e_b^j\right)\left(\mu_c^k-\nabla_c 
e_0^k-\epsilon^{klm}e_{cl}A_{0m}\right)-\zeta\nabla_b e_{ci}-
\epsilon_{ijk}e_{0}^{j}\nabla_be_c^k\right]=0\label{20}\\
& & \hspace{-2cm}
{\tilde\eta}^{abc}\left[\left(\frac{1}{2}\delta_{ik}
\nabla_b\phi-\epsilon_{ijk}e_b^j\right)\left(\lambda_c^k-\nabla_c A_0^k\right)-\frac{1}{2}\zeta 
F_{bci}+\frac{1}{2}\epsilon_{ijk}e_0^jF_{bc}^k\right]=0\label{21}\\
& & \nonumber\\
& & \hspace{-2cm}
{\tilde\eta}^{abc}\left[\left(\mu_a^i-\nabla_a e_0^i-
\epsilon^{ijk}e_{aj}A_{0k}\right)F_{bci}+2\left(\lambda_a^i-\nabla_a A_0^i\right)\nabla_b 
e_{ci}\right]=0\label{22}
\end{eqnarray}
The conservation in time of (\ref{18}) and (\ref{19}) does not generate new secondary constraints 
but only the following conditions on the Lagrange multipliers
\begin{eqnarray}
& & \hspace{-.5cm}
{\tilde\eta}^{abc}\left\{\nabla_a\left[\left(\frac{1}{2}
\delta_{ik}\nabla_b\phi+
\epsilon_{ijk}e_{b}^j\right)\mu_c^k\right]-(\nabla_a\zeta)
(\nabla_b e_{ci})-
\right.\label{23}\\
& & \hspace{6cm}\left.-\left(\frac{1}{2}\delta_{ik}\nabla_a\phi+\epsilon_{ijk}e_a^j
\right)\epsilon^k_{\;\;lm}\lambda_b^l e_c^m\right\}=0\nonumber\\
& & \hspace{-.5cm}{\tilde\eta}^{abc}\left\{\left(\frac{1}{2}\delta_{ik}
\nabla_a\phi-\epsilon_{ijk}e_{a}^j\right)\nabla_b\lambda_c^k-
\frac{1}{2}\epsilon_{ijk}\mu_a^jF_{bc}^{k}+\frac{1}{2} F_{ab}^i\nabla_c\zeta\right\}=0\label{24}
\end{eqnarray}
In principle, one expects that some combination of the second class constraints will be first 
class. The way to find out if this is the case is to solve the equations for the Lagrange 
multipliers. As we show in the appendix it is possible to find $\mu_a^i$ from (\ref{20}) and 
$\lambda_a^i$ from (\ref{21}) and write them in terms of $\zeta$, $e_a^i$, $e_0^i$, $A_a^i$, and 
$A_0^i$
\begin{eqnarray}
& & \mu_a^i=\nabla_a e_0^i+\epsilon^{ijk}e_{aj}A_{0k}+\PP_{a\;\;b}^{\;\;i\;\;j}
{\tilde\eta}^{bcd}\left(\zeta\nabla_c e_d^j+\epsilon_{jkl}e_{0}^{k}\nabla_c 
e_d^l\right)\label{25a}\\
& & \lambda_a^i=\nabla_a A_0^i-\frac{1}{2}\PP_{b\;\;a}^{\;\;j\;\;i}{\tilde\eta}^{bcd}
\left(\zeta F_{cdj}-\epsilon_{jkl}e_{0}^{k}F_{cd}^l\right)\label{25b}
\end{eqnarray}
where $\PP_{a\;\;b}^{\;\;i\;\;j}$ (which is calculated in the appendix) satisfies ${\tilde 
P}^{a\;\;b}_{\;\;i\;\;j}\PP_{b\;\;c}^{\;\;j\;\;k}=\delta^a_c
\delta^i_k$.
We have made the ansatz that the triad is non-degenerate (and we will continue to do so 
throughout the paper). After some tedious algebra it is possible to verify that (\ref{22}-
\ref{24}) are identically satisfied by the previous $\mu_a^i$ and $\lambda_a^i$. We leave (some) 
of the details for the appendix.

We want to stress here the importance of paying attention to the conditions on the Lagrange 
multipliers that appear in the Hamiltonian analysis. If one knows beforehand what a theory means, 
one can usually skip the arduous solution of the consistency equations as  one does not need to 
know the explicit form of the Lagrange multipliers once all the first class constraints have been 
identified. However, it is true, in general, that the Lagrange multiplier equations themselves 
may imply additional constraints (they are non-homogeneous linear equations) so, if one does not 
know the meaning of the theory one is dealing with, great attention must be paid to these 
equations in order to avoid missing some of the constraints and completely fail in the 
interpretation of the theory.

Substituting (\ref{25a}-\ref{25b}) in $H_T$ we get
$$H_T=\int_{\cal M} d^3x\;\left\{e_0^i\left[{\tilde\eta}^{abc}\left(\frac{1}{2}
 F_{abi}\nabla_c \phi-\epsilon_{ijk}e_a^j F_{bc}^k\right)-\nabla_a{\tilde\sigma}^a_i-
\epsilon_{ijk}{\tilde\sigma}^a_l
\PP_{a\;\;b}^{\;\;l\;\;j}{\tilde\eta}^{bcd}\nabla_c e_{d}^{k}-\right.\right.$$
\vspace*{-1.3cm}
\begin{eqnarray}
& & \hspace{1cm}\left.-\frac{1}{2}\left({\tilde\pi}_j^a+
{\tilde\eta}^{abc}( e_{bj}\nabla_c\phi-\epsilon_{jkl}e_b^k e_c^l)\right)\PP_{d\;\;a}^{\;\;m\;\;j}
{\tilde\eta}^{def}\epsilon_{imn}F_{ef}^{n}\right]-\nonumber\\
& & \hspace{1cm}-A_0^i\left(\nabla_a{\tilde\pi}_i^a+
\epsilon_{ijk}e_a^j{\tilde\sigma}^{ak}\right)+\lambda_i
{\tilde\pi}^i+\mu_i{\tilde\sigma}^i+\label{26}\\
& & \hspace{1cm}+\zeta\left[2{\tilde p}-{\tilde\eta}^{abc}e_a^iF_{bci}+{\tilde\sigma}^a_i
\PP_{a\;\;b}^{\;\;i\;\;j}{\tilde\eta}^{bcd}\nabla_ce_{dj}-
\right.\nonumber\\
& & \hspace{1cm}-\left.\left.\frac{1}{2}\PP_{b\;\;a}^{\;\;j\;\;i}
{\tilde\eta}^{bcd}
F_{cdj}\left({\tilde\pi}_i^a+{\tilde\eta}^{aef}( e_{ei}\nabla_f\phi-\epsilon_{ikl}e_e^k 
e_f^l)\right)\right]\right\}\nonumber
\end{eqnarray}
The terms proportional to $e_0^i$ and $A_0^i$ together give a first class Hamiltonian and the 
terms proportional to $\zeta$, $\lambda_i$, and $\mu_i$ are first class constraints (each of 
them). Of course, we have also all the remaining constraints provided by (\ref{15a}-\ref{15b}, 
\ref{18}, \ref{19}). The first class constraints ${\tilde\pi}^i=0$ and ${\tilde\sigma}^i=0$ imply 
that $A_0^i$ and $e_0^i$ are arbitrary functions so we can just remove ${\tilde\pi}^i=0$ and 
${\tilde\sigma}^i=0$ from (\ref{26}). Furthermore, as now $A_0^i$, $e_0^i$, and $\zeta$
are arbitrary and $H_T$ is first class, the expressions that they multiply (linear combinations 
of first and second class constraints) must be first class constraints. In this way we get three 
sets of first class constraints plus the following independent second class constraints
\newpage
\begin{eqnarray}
& & {\tilde\pi}^a_i+{\tilde\eta}^{abc}\left[e_{bi}\nabla_c\phi-
\epsilon_{ijk}e_b^je_c^k\right]=0\label{27}\\
& & {\tilde\sigma}^a_i=0\label{28}
\end{eqnarray}
These are very easy to deal with. In practice it is enough to remove ${\tilde\sigma}^a_i$ from 
the first class constraints  and write $e_a^i$ in terms of $\phi$ and ${\tilde\pi}^a_i$ by 
solving (\ref{27})
$$
e_a^i=\frac{1}{4{\tilde{\tilde\pi}}}\eut_{abc}\left\{\pm
\left[2{\tilde{\tilde\pi}}-({\tilde\pi}^d_l\nabla_d\phi )^2\right]^{1/2}\epsilon^{ijk}
{\tilde\pi}^b_j {\tilde\pi}^c_k-2
({\tilde\pi}^d_k\nabla_d\phi){\tilde\pi}^{bk}{\tilde\pi}^{ci}
\right\}
$$
where ${\tilde{\tilde\pi}}\equiv\det {\tilde\pi}^a_i$.
The final Hamiltonian description is very simple. The phase space is coordinatized by the 
canonically conjugate pairs ($A_a^i$, ${\tilde\pi}^a_i$) and ($\phi$, ${\tilde 
p}$)\footnote{\noindent This is the symplectic structure given by the Dirac brackets.} and the 
first class constraints are
\begin{eqnarray}
& & \nabla_a{\tilde\pi}^a_i=0\nonumber\\
& & {\tilde\pi}^b_i F_{ab}^i+{\tilde p}\nabla_a \phi=0\label{30}\\
& & {\tilde p}\mp\frac{1}{2}\left[2{\tilde{\tilde\pi}}-({\tilde\pi}^d_l\nabla_d\phi )^2\right]^{-
1/2}\epsilon^{ijk}
{\tilde\pi}^a_i{\tilde\pi}^b_jF_{abk}=0\nonumber
\end{eqnarray}
They are the Gauss law, that generates $SO(3)$ gauge transformations, the vector constraint that 
(essentially) generates diffeomorphisms, and a scalar constraint linear in ${\tilde p}$. They are 
first class constraints. It is convenient to write them in ``weighted" form
\begin{eqnarray}
& & G(N^i)=\int_{\Sigma}d^3x\;N^i\nabla_a{\tilde\pi}^a_i
\nonumber\\
& & V(N^a)=\int_{\Sigma}d^3x\;N^a\left({\tilde\pi}^b_i
F_{ab}^i+{\tilde p}\nabla_a\phi\right)\label{31}\\
& & S(N)=\int_{\Sigma}d^3x\;N\left\{{\tilde p}\mp\frac{1}{2}\left[2{\tilde{\tilde\pi}}-
({\tilde\pi}^d_l\nabla_d\phi )^2\right]
^{-1/2}\epsilon^{ijk}{\tilde\pi}^a_i{\tilde\pi}^b_jF_{abk}
\right\}\nonumber
\end{eqnarray}
The three-dimensional diffeomorphisms are generated by the combination of the Gauss law and the 
vector constraint $D(N^a)\equiv G(N^a A_a^i)-V(N^a)$.
We can write now the constraint algebra
\begin{eqnarray}
& & \left\{G(N^i),G(M^i)\right\}=G([N,M]^i)\hspace{1.5cm}
{\rm with}\hspace{1.5cm}[N,M]^i\equiv\epsilon^{ijk}N_jM_k
\nonumber\\
& & \left\{G(N^i),V(M^a)\right\}=0\nonumber\\
& & \left\{G(N^i),S(M)\right\}=0\label{32}\\
& & \left\{D(N^a),D(M^b)\right\}=D(-[N,M]^a)\hspace{.5cm}
{\rm with}\hspace{.5cm}[N,M]^a\equiv N^b\partial_b M^a-M^b\partial_b N^a
\nonumber\\
& & \left\{D(N^a),S(M)\right\}=S(-N^a\nabla_a M)\nonumber\\
& & \left\{S(N),S(M)\right\}=V\left[(N\partial_aM-
M\partial_aN)\frac{4{\tilde\pi}^a_i{\tilde\pi}^{bi}}
{2{\tilde{\tilde\pi}}-({\tilde\pi}^d_l\nabla_d\phi )^2}\right]\nonumber
\end{eqnarray}
Several remarks are now in order. First, we see that the constraints are first class. As we have 
20 canonical variables per space point in $\Sigma$ and seven first class constraints we have 
three degrees of freedom per space point --one more that in G.R.--. Second, the Poisson bracket 
of the scalar constraint with itself closes and gives the vector constraint. This is in agreement 
with what one would expect from the arguments given by Hojman, Kucha\v{r} and Teitelboim in 
\cite{HKT} where they showed that the algebra of space-time deformations implied a constraint 
algebra of the type given by (\ref{32}). Third, the structure of the scalar constraint is quite 
suggestive; it has two terms, one linear in ${\tilde p}$ and another proportional to the scalar 
constraint in the Euclidean Ashtekar  formulation for G.R. This may signal a previously unnoticed 
relation between the Husain-Kucha\v{r} model and G.R.

From (\ref{30}-\ref{31}) we can interpret the model very easily. It is enough to impose the gauge 
fixing condition $\phi=0$ (admissible because $\left\{\phi(x), {\tilde 
p}(y)\right\}=\delta^3(x,y)$) to get rid of the scalar constraint and the $\phi$ dependent part 
of the vector constraint to recover the constraints of the usual H-K model, namely
\begin{equation}
\begin{array}{c}
\nabla_a{\tilde\pi}^a_i=0\\
\\
{\tilde\pi}^a_i F_{ab}^i=0
\end{array}
\label{33}
\end{equation}
This means that in our formulation of the model the gauge
orbits have one extra dimension so, in rigor, the models are equivalent only modulo gauge 
transformations.
At this point the reader may have the temptation to think that, after all, it is trivial to add a 
scalar constraint
to (\ref{33}) in order to have a time variable (just take ${\tilde p}=0$ and add the term 
necessary to generate diffeomorphisms on $\phi$ and ${\tilde p}$ to the vector constraint). The 
formulation thus obtained is, obviously, equivalent to ours (and can be derived from the action 
(\ref{9}) by removing the derivatives of $\phi$ with an integration by parts). However, it is 
much less obvious (and less trivial) the fact that with a suitable choice of a scalar constraint 
one gets, not only a time variable, but also a way to interpret the H.K model as a theory for 
non-degenerate four-metrics at the Lagrangian level.

\section{The four-dimensional Picture: Non-degenerate 4-metrics}

The four dimensional field equations coming from the action (\ref{9}) are
\begin{eqnarray}
& & {\tilde\eta}^{abcd}\left\{F_{abi}\nabla_c\phi-
2\epsilon_{ijk}e_{a}^{j}F_{bc}^{k}\right\}=0\label{34a}\\
& & {\tilde\eta}^{abcd}\left\{\nabla_{a}e_{bi}\nabla_c\phi+
2\epsilon_{ijk}e_{a}^{j}\nabla_{b}e_{c}^{k}\right\}=0
\label{34b}\\
& & {\tilde\eta}^{abcd}(\nabla_a e_{b}^{i})F_{cdi}=0\label{34c}
\end{eqnarray}
If we have a solution to these equations we can build a four-metric from the tetrad given by 
$(\nabla_a\phi, e_a^i)$ as $g_{ab}=\pm\nabla_a\phi\nabla_b\phi+e_a^i e_{bi}$. Notice that it is 
possible to write both Euclidean and Lorentzian 4-metrics by choosing the sign in front of the 
$\nabla_a\phi\nabla_b\phi$ term. In general one expects that 
$g_{ab}$ is non-degenerate as can be checked by simply showing some solutions to (\ref{34a}-
\ref{34c}) such as
\begin{equation}
A_a^i=0,\hspace{2cm}\phi=x^0\hspace{2cm}e_a^i=\left[
\begin{array}{ccc}
0 & 0 & 0\\
1 & 0 & 0\\
0 & 1 & 0\\
0 & 0 & 1
\end{array}\right]
\label{35}
\end{equation}
As it can be seen (\ref{35}) provides both the Euclidean and the Minkowski metric in ${\rm 
I\!R^4}$. We see that we can solve the (apparent) paradox presented in the introduction by using 
the scalar field that is present now in the field equations to build non-degenerate four-metrics.

It is interesting at this point to compare the new equations (\ref{34a}-\ref{34c}) with the old 
ones (\ref{2}). For starters we seem to have one more equation now that we had before; however, 
as we show below, this equation is not independent of the others and, also, any solution to 
(\ref{2}) is a solution to it. In the following we use a procedure similar to the one that 
appears in  section III of \cite{HuKu}. Let us write
$$
\begin{array}{l}
E_{ab}^i\equiv\nabla_{[a}e_{b]}^i\\
\\
{\tilde n}^a
\equiv\frac{1}{3!}{\tilde\eta}^{abcd}\epsilon_{ijk}e_b^i e_c^j e_d^k\\
\\
{\tilde\eta}^a_i
\equiv-\frac{1}{2}{\tilde\eta}^{abcd}\epsilon_{ijk} e_b^j e_c^k\nabla_d\phi
\end{array}
$$
Now ${\tilde n}^a{\tilde n}^b E_{ab}^i=0$ implies that there must exist ${\tilde E}_j^{\;\;i}$ 
such that 
$$
{\tilde n}^a E_{ab}^i=e_b^j{\tilde E}_j^{\;\;i}
$$
Notice that $e_a^i$ satisfy ${\tilde n}^a {\tilde E}_a^i=0$ so that any linear combination of the 
$e_a^i$ such as $e_a^j{\tilde E}_j^{\;\;i}$ will also satisfy ${\tilde n}^a{\tilde E}_j^{\;\;i} 
e_a^j=0$. By the same reasoning there must exist ${\tilde F}_j^{\;\;i}$ such that 
$$
{\tilde n}^a F_{ab}^i=e_b^j{\tilde F}_j^{\;\;i}
$$
We define also (we suppose ${\tilde n}^d\nabla_d\phi\neq 0$)$$
e^{kl}\equiv\frac{1}{({\tilde n}^d\nabla_d\phi)^2}\epsilon^{ijk}{\tilde\eta}^a_i
{\tilde\eta}^b_jE_{ab}^l,\hspace{1cm}
f^{kl}\equiv\frac{1}{({\tilde n}^d\nabla_d\phi)^2}\epsilon^{ijk}{\tilde\eta}^a_i
{\tilde\eta}^b_jF_{ab}^l
$$
We can  extract all the content from the equations (\ref{34a}-\ref{34c}) by multiplying the first 
two by 
$$
\epsilon^{ijk}{\tilde\eta}^a_i{\tilde\eta}^b_j
{\tilde\eta}^c_k,\hspace{2cm} \epsilon^{ijk}{\tilde n}^{[a}{\tilde\eta}^b_j
{\tilde\eta}^{c]}_k
$$
and the scalar equation (\ref{34c}) by 
$$
\epsilon^{ijk}{\tilde n}^{[a}
{\tilde\eta}^b_i{\tilde\eta}^{c}_j
{\tilde\eta}^{d]}_k
$$
(which is proportional to ${\tilde\eta}^{abcd}$).

The result that we obtain from (\ref{34a}) is that $F^{ij}=-\frac{1}{2}\left(f^{ij}-
\frac{1}{2}\delta^{ij}f\right)$ and $f^{ij}$ is symmetric and from (\ref{34b}) that 
$E^{ij}=\frac{1}{2}\left(e^{ij}-\frac{1}{2}\delta^{ij}e\right)$ and $e^{ij}$ is symmetric, where 
$e$ and $f$ are the traces of $e_{ij}$ and $f_{ij}$ respectively. In terms of $e^{ij}$, $f^{ij}$, 
$E^{ij}$, and $F^{ij}$ the scalar equation (\ref{34c}) gives $e^{ij}F_{ij}+f^{ij}E_{ij}=0$; we 
see now that all the solutions to (\ref{34a}) and (\ref{34b}) are solutions to the  scalar 
equation and, hence, it is redundant.

If we consider now the standard H-K equations we see that there is no scalar equation there. It 
is possible to extract the content of (\ref{2}) by using the procedure introduced above. The only 
difference now is that we need an auxiliary scalar function (for example the one that gives the 
foliation used in the passage to the Hamiltonian formulation) to define ${\tilde\eta}^a_i$. We 
immediately find the result that appears in \cite{HuKu} $e^{[ij]}=0$, $f^{[ij]}=0$, $E^{ij}=0$, 
and $F^{ij}=0$,  so that now it is also true that the scalar equation (\ref{34c}) is satisfied. 
In order to compare the solutions to (\ref{2}) and (\ref{34a}-\ref{34c}) one must take into 
account the new symmety present in the model due to the introduction of $\phi$.

\section{From the Old to the New Husain-Kucha\v{r} Model: Equivalence at the Lagrangian Level.}

Although we have seen from the Hamiltonian analysis that the new and all formulations of the H-K 
model are strictly equivalent it is instructive to understand this from an independent point of 
view because the actions (\ref{1}) and (\ref{9}) look quite different (in fact one could claim 
that (\ref{9}) is really ``closer" to the self-dual action for G.R. than to the H-K action).

The key idea to show this equivalence is the last result of the previous section, i.e. the fact 
that every solution to the ordinary H-K equations (\ref{2}) also satisfies (\ref{34c}). This 
means that nothing changes if we add this condition to the action (\ref{1}) with a scalar 
Lagrange multiplier $\phi$. In this way we get
$$
{\hat S}_1=\int_{\cal M}d^4 x\,{\tilde\eta}^{abcd}\left[
-\phi F_{ab}^i\nabla_c e_{di}+\epsilon_{ijk}e_a^i e_b^j F_{cd}^k\right]
$$
which is obviously equivalent to (\ref{9}). Actually we can go even further. From the H-K 
equations it is straightforward to show that (\ref{2}) implies
$$
{\tilde\eta}^{abcd}\nabla_a(\epsilon_{ijk}e_b^ie_c^j e_d^k)=0
$$
so that even the action\footnote{\noindent There are even more possibilities that we do not 
discuss here.}
$$
{\hat S}_2=\int_{\cal M}d^4 x\,{\tilde\eta}^{abcd}\left[
\phi F_{ab}î\nabla_c e_{di}+\epsilon_{ijk}e_a^i e_b^j F_{cd}^k+\psi \epsilon_{ijk}\nabla_a (e_b^i 
e_c^j e_d^k)\right]
$$
describes the H-K model. This last action admits an interesting interpretation. If we choose 
$\psi(x)=-\frac{\Lambda}{3!}\phi(x)$ with $\Lambda$ a real constant and consider the tetrad 
$e_a^I\equiv(\nabla_a\phi, e_a^i)$ whose inverse is given by
$$
e^a_I\equiv\frac{1}{\det e_a^I}\left[
\begin{array}{c}
{\tilde n}^a\\
{\tilde \eta}^a_i
\end{array}
\right]
$$
with ${\tilde n}^a$ and ${\tilde\eta}^a$ as defined in the previous section we see that the added 
term is, in fact
$$
\int_{\cal M}d^4 x\,\Lambda(\det e_a^I)
$$
that is, a cosmological constant term. This is the simplest (and trivial) instance of a matter 
coupling to the H-K model using the, now available, non-degenerate four metric.

An additional curious fact is that the previous term is equivalent to
$$
\int_{\cal M}d^4x\,\Lambda (\det e_a^I)e^a_J e^{aJ}\nabla_a\phi\nabla_b\phi
$$
i.e. the coupling of the $\phi$ field to the ``non-degenerate H-K model" as a free scalar. A 
Hamiltonian analysis of these last actions with a ``cosmological constant" following the lines of 
section IV shows their equivalence with the usual H-K model.

\section{Conclusions and Perspectives}

As we have shown in the paper it is possible to describe the Husain-Kucha\v{r} model with an 
action principle for non-degenerate metrics. We have accomplished this by introducing a scalar 
field in such a way that adds no new degrees of freedom. This scalar plays, in a sense, the role 
of a time variable not only because we have now a Hamiltonian constraint that is linear in its 
canonical momentum but also  because it allows dynamics to be referred to it. Our proposal should 
be compared to those of other authors (especially \cite{Hu} and \cite{Rov1}). In these papers a 
scalar field is included as a means to define quantum gauge invariant observables, quoting 
Rovelli ``matter observables which can be used to dynamically determine surfaces, the areas of 
which, we can measure". Our contribution in this respect is that we have managed to achieve this 
goal without introducing new degrees of freedom in the model. We find it quite appealing that  in 
this process we get a nice interpretation of the scalar $\phi$ as time. Not only can we do this 
but also, as a side result, we have now the possibility of coupling ordinary matter to the model. 
This provides a type of theories that lie in between those that have a matter evolving in a non-
dynamical background and full G.R. We think that a lot can be learned from looking at these 
theories; we plan to study them in the future. Notice, by the way, that we have the choice of 
coupling the matter fields to Euclidean or Lorentzian metrics, depending on the choice of the 
sign in the first term of the four-metric $g_{ab}=\pm \nabla_a\phi \nabla_b\phi+e_a^i e_{bi}$. 

We want to remark at this point that not knowing beforehand what the meaning of the action 
(\ref{9}) is, one should be very careful in order to avoid missing constraints crucial for the 
interpretation of the theory. That is why we have paid so much attention to the solution of the 
equations for the Lagrange multipliers. Also, we emphasize again the contradiction in claiming 
that the Husain-Kucha\v{r} model only allows for the existence of degenerate four-metrics whereas 
it is obviously an extension of both Euclidean and Lorentzian G.R. We believe that we have 
clearly solved this seemingly paradoxical fact in the paper.

\section{Appendix}

As we have said in the main text of the paper we have paid special attention to the solution of 
the Lagrange multiplier equations (\ref{20}-\ref{24}). The strategy that we have followed is 
simple. First solve (\ref{20}) for $\mu_a^i$ and (\ref{21}) for $\lambda_a^i$ and plug the result 
in the remaining ones. The result that we have obtained (for non-degenerate triads) shows that 
once we write these Lagrange multipliers in terms of $\zeta$, $A_{0i}$, $e_{0i}$, 
$A_{ai}$, $e_{ai}$, and $\phi$ the remaining equations are identically satisfied.

In order to solve (\ref{20}) and (\ref{21}) we need to compute the inverses (that we denote 
$\PP_{a\;\;b}^{\;\;i\;\;j}$) of the $9\times 9$ matrices
$$
{\tilde P}^{\;a\;\;c}_{\;\;\;i\;\;k}(x)
\equiv{\tilde\eta}^{abc}\left(
\frac{1}{2}\delta_{ik}\nabla_b\phi+\epsilon_{ijk}e_b^j\right)
$$
and its transponse
$$
{\tilde P}^{\;c\;\;a}_{\;\;\;k\;\;i}(x)
\equiv{\tilde\eta}^{abc}\left(-
\frac{1}{2}\delta_{ik}\nabla_b\phi+\epsilon_{ijk}e_b^j\right)
$$
where $^a_{\;\;i}$ are ``double indices" that take the nine different values that make these 
matrices $9\times 9$. The best way to build their inverses is to explicitly solve the equation
\begin{equation}
{\tilde M}^{\;a\;\;b}_{\;\;\;i\;\;j}X_b^j\equiv{\tilde\eta}^{abc}
\left(\delta_{ij}v_c+\epsilon_{ijk}e_c^k\right)X_b^j={\tilde J}^a_i
\label{a1}
\end{equation}
First we introduce the inverse triad $e^a_i$ such that $e^a_i e_a^j=\delta_i^j$ and write 
${\tilde\eta}^{abc}={\tilde e}\epsilon^{ijk}e^a_ie^b_j e^c_k$ (here ${\tilde e}$ is the non-zero 
determinant of the triad). Introducing this in (\ref{a1}), expanding, and using the notation
$$
X_{ij}\equiv e^a_i X_{aj},\hspace{5mm}X\equiv e^a_i X_{a}^{i}\hspace{5mm} J^a_i\equiv{\tilde 
J}^a_i/{\tilde e},\hspace{5mm}J_{ij}\equiv e_{ai} J^a_{j},\hspace{5mm}, 
J\equiv e_{ai} J^{ai}
$$
we get
$$
\epsilon^{lmn}e^a_l X_{mi}v_n+\epsilon^{lmk}\epsilon_{ijk}e^a_l X_{m}^{\;\;j}=J^a_i
$$
which, after multiplying by $e_{al}$ transforms into an equation that only involves objects with 
internal indices.
\begin{equation}
\epsilon_l^{\;\;mn}X_{mi}v_n+X \delta_{il}-X_{il}=j_{li}
\label{a4}
\end{equation}
Let us now take the trace of (\ref{a4}), multiply it by $\epsilon_{ilp}v^p$ and by $v_i v_l$. We 
find the following three equations
\begin{eqnarray}
& &-\epsilon^{ijk}X_{ij} v_k+2X=J\label{a5a}\\
& & X_{ij}v^iv^j-v^2X-\epsilon^{ijk}X_{ij}v_k=-\epsilon^{ijk}J_{ij}v_k\label{a5b}\\
& & X v^2-X_{ij}v^iv^j=J_{ij}v^iv^j\label{a5c}
\end{eqnarray}
where $v^2\equiv v_iv^i$. Adding (\ref{a5b}) and (\ref{a5c}) and using (\ref{a5a}) gives
$$
X=\frac{1}{2}\left[J+\epsilon^{ijk}J_{ij}v_k-v^iv^jJ_{ij}\right]
$$
This means that we know how to express $X$ in (\ref{a4}) in terms of $J_{ij}$ and $v_i$. If we 
look now at how the indices in the remaining $X_{ij}$ appear we see that the $i$ index is at both 
the second and the first place. If we could find the way to have both $i$ indices at the second 
place the remaining equation would be very easy to solve by inverting a simple $3\times 3$ 
matrix. To this end we need to know the expression for $X_{[ij]}$ in terms of $J_{ij}$. This can 
be computed by multiplying (\ref{a1}) both by $v_a$ and $\epsilon_{ilm}e_a^m$ and eliminating 
tangent space indices as before. One gets
$$
X_{[ij]}=J_{[ij]}-\frac{1}{2}\epsilon_{ijk}J^{lk}v_l
$$
Using this result back in (\ref{a4}) we have
\begin{equation}
X^{k}_{\;\;i}(\delta_{jk}-\epsilon_{jkl}v^l)=\frac{1}{2}\delta_{ij}(J+\epsilon^{pqr}
J_{pq}v_r-v^pv^qJ_{pq})+\epsilon_{ijk}J^{lk}v_l-J_{ij}
\label{a8}
\end{equation}
Multiplying (\ref{a8}) by 
$$
\frac{1}{1+v^2}\left[\delta_{nj}+v_nv_j+\epsilon_{njs}v^s
\right]
$$
and reintroducing the triads we finally get
\begin{eqnarray}
& & \MM_{a\;\;b}^{\;\;i\;\;j}=
\frac{1}{2{\tilde e}(1+v^2)}\left[\delta_{nl}+v_nv_l+
\epsilon_{nlr}v^r\right]\times\nonumber\\
& & \times\left[\delta^{il}\delta_{k}^{j}-
2\delta^{i}_{k}\delta^{jl}+\delta^{il}\epsilon^j_{\;\;mk}v^m+2\epsilon^{ilj}v_k-
v_kv^j\delta^{il}\right]e_a^n e_b^k
\label{a10}
\end{eqnarray}
The inverses of ${\tilde P}^{\;a\;\;b\;\;}_{\;\;i\;\;j}$ and its transponse are immediately 
obtained from the (\ref{a10}). With them it is possible to check by direct substitution that the 
consistency equations (\ref{22}-\ref{24}) are identically satisfied. In practice the best 
strategy to do this is the following. First, eliminate the tangent space indices by multiplying 
by suitable combinations of inverse triads, then use the constraints (\ref{18}) and (\ref{19}) in 
the form
$$
\left[-\frac{1}{2}\delta_{ik}\nabla_a\phi+\epsilon_{ijk}e_a^j\right]
{\tilde\eta}^{abc}\nabla_b e_{ci}=0
$$
$$ \left[+\frac{1}{2}\delta_{ik}\nabla_a\phi+\epsilon_{ijk}e_a^j\right]
{\tilde\eta}^{abc}F_{bci}=0
$$
In order to check (\ref{23}) it is very useful to use the following identity
$$
{\tilde\eta}^{abc}\left(P_{ai}^k\epsilon_{klm}-{\hat 
P}_{akl}\epsilon^k_{\;\;mi}\right)e_{c}^{m}\Lambda_{b}^{l}=0
$$
where
$$
P_{aik}\equiv v_a\delta_{ik}-\epsilon_{ijk}e_{a}^{j}
$$
$$ {\hat P}_{aik}\equiv v_a\delta_{ik}+\epsilon_{ijk}e_{a}^{j}
$$
and $\Lambda_b^l$ is arbitrary.

\noindent{\bf Acknowledgments}

The authors want to thank our colleagues G. Immirzi, J. Julve, J. Leon and G. Mena by their 
useful comments on this paper. J.F.B.G. wants to thank also J. M. Mart\'{\i}n Garc\'{\i}a for a 
very enlightening discussion. J.F.B.G. and R.T. are supported by C.S.I.C. contracts.

\end{document}